\documentclass[
11pt]{elsarticle}

\usepackage{amssymb}

\usepackage[a4paper,left=3.9cm, right=3.9cm,top=4.3cm,bottom=4.7cm]{geometry}

\usepackage[latin1]{inputenc}
\usepackage[boxed]{algorithm2e}
\usepackage{amsmath}\usepackage{amsthm}\usepackage{amssymb}\usepackage{mathrsfs}\usepackage{amscd}\usepackage{graphicx}\usepackage[FIGTOPCAP]{subfigure}\usepackage{amsfonts}\usepackage{amsxtra}\usepackage{color}
\usepackage{amscd}
\usepackage[subfigure]{tocloft}

\DeclareMathOperator{\PFP}{PFP}

\DeclareMathOperator{\trace}{trace}

\DeclareMathOperator{\R}{\mathbb{R}}

\theoremstyle{definition}
\theoremstyle{plain}

\journal{arXiv.org}

\begin{document}
\newcommand{\patterning}{patterning of granular rods}

\begin{frontmatter}



\title{Frame theory in directional statistics
}


\author[a,c]{Martin Ehler\corref{cor1}}
\ead{ehlermar@mail.nih.gov}
\cortext[cor1]{Corresponding author}

\author[a,b]{Jennifer Galanis}
\ead{galanis@fh.huji.ac.il}
\address[a]{National Institutes of Health, National Institute of Child Health and Human Development, Section on Medical Biophysics, 
Bethesda, MD 20892}
\address[c]{University of Maryland, Department of Mathematics, Norbert Wiener Center,
College Park, MD 20742}
\address[b]{Institute of Chemistry and The Fritz Haber Center, The Hebrew
University of Jerusalem, Jerusalem, 91904, Israel.}

\begin{abstract}
Distinguishing between uniform and non-uniform sample distributions is a common problem in directional data analysis; however for many tests, non-uniform distributions exist that fail uniformity rejection. By merging directional statistics with frame theory, 
we find that probabilistic tight frames yield non-uniform distributions that minimize directional potentials, leading to failure of uniformity rejection for the Bingham test. Finally, we apply our results to model patterns found in granular rod experiments. 
%
\end{abstract}


\end{frontmatter}


\section{Introduction}
Observations that inherit a direction occur in many scientific disciplines. For example, directional data arise naturally in the biomedical field for protein structure, cell-cycle, and circadian clock experiments \cite{Boomsma:2008aa,Chasse:1988aa,Levine:2002aa,Mardia:2008aa,Mardia:2007aa,Rueda:2009aa}. Further examples occur in statistical mechanics, where experiments containing only rod-shaped particles can develop complex directional ordering \cite{Galanis:2006aa,Galanis:2010ab,Gennes:1993aa,Onsager:1949aa}. A simple pattern change for rod shaped objects is the density-dependent, order-disorder phase transition \cite{Gennes:1993aa,Onsager:1949aa} shown for macroscopic granular rod experiments \cite{Galanis:2006aa} in Fig.~1(a) and 1(b). Quantification of this transition relies on a principle component analysis (PCA) type measure that is linked with statistical mechanical theories \cite{Gennes:1993aa}. When applied to experimental samples whose rod orientations shift from uniform to unidirectional, this measure finds the dominant direction (director) and strength of rod ordering (order parameter).


In reality, however, experimental rod orientations are rarely
unidirectional, and spatial distortions in the director field frequently
occur. These distortions may result from fluctuations and/or other
competing forces, like those exerted by the container boundaries in
Fig.~1(c) and 1(d). Accurately quantifying rod orientations can, in
fact, yield information about the collective behavior of rods, for
example elastic properties \cite{Galanis:2010ab,Gennes:1993aa}. While accurate orientation measurements of
molecular-sized rods require special techniques
\cite{Hudson:1989aa}, recent advances in single molecule detection may make such measurements more widely accessible \cite{Chang:2010aa,Xiao:2010aa}. For example, ``labeled'' rods can be inserted into various environments and serve as local directional sensors by aligning with the rod-shaped material around them. This technique can in principle be used in environments as complicated as a cell \cite{Chang:2010aa,Xiao:2010aa} and potentially uncover intricate patterns that require more sophisticated measures of directional order.


The resulting complex patterning may reduce the value of the order parameter, sometimes to the point that the sample is inaccurately classified as disordered, Fig.~1(d). To predict which multidirectional patterns cause such misclassifications, we merge directional statistics with frame theory. 
Frames have proven useful in fields like spherical codes, compressed sensing, signal processing, and wavelet analysis \cite{Casazza:2003aa,Christensen:2003aa,Daubechies:1986aa,Ehler:2007aa,Ehler:ab,Ehler:2010ae,Ehler:2008ab,Ehler:aa,Feichtinger:2003aa,Grochenig:2001aa}. A frame is a basis-like system that spans a vector space but allows for linear dependency, which can be used to reduce noise, find sparse representations, or obtain other desirable features unavailable with orthonormal bases. Tight frames even provide a parseval type formula similar to orthonormal bases. Moreover, the frame concept has recently been generalized to probability distributions on the unit sphere \cite{Ehler:2010aa}.    

To analyze granular rod patterning, we consider statistical testing for directional uniformity, focusing on the Bingham test. We characterize non-uniform sample distributions that lead to failure of rejection and find that these distributions are probabilistic tight frames. Since these frames are well-understood in terms of algebraic/geometric conditions \cite{Ehler:2010aa}, further synergistic effects may develop between directional statistics and frame theory.

\section{Directional statistics}\label{section:directional statistics}
Common tests in directional statistics focus on whether or not a sample on the unit sphere $S^{d-1}=\{x\in\R^d : \|x\|=1\}$ is uniformly distributed. Here, we concentrate on two elementary tests for uniformity, \emph{Rayleigh} and \emph{Bingham}. Given a discrete sample $\{x_i\}_{i=1}^n\subset S^{d-1}$, we follow the textbook \cite{Mardia:2008aa} and define the \emph{mean} as
\begin{equation}\label{eq:mean}
\bar{x} = \frac{1}{n}\sum_{i=1}^n x_i,
\end{equation}
where the polar representation $\bar{x}=\bar{r}\bar{x}_0$ splits the mean into a \emph{mean direction} $\bar{x}_0\in S^{d-1}$ and a \emph{mean resultant length} $\bar{r}=\|\bar{x}\|$. The \emph{Rayleigh test} rejects the hypothesis of uniformity if $\bar{r}$ is large. More precisely, the asymptotic large-sample distribution of $dn\bar{r}^2$ under uniformity is $\chi^2_d$ distributed with an error $\mathcal{O}(n^{-1})$, while the modified Rayleigh statistic $(1-\frac{1}{2n})dn\bar{r}^2+\frac{1}{2n(d+2)}d^2n^2\bar{r}^4$ is $\chi^2_d$ distributed with an error $\mathcal{O}(n^{-2})$ \cite{Mardia:2008aa}.

To describe the \emph{Bingham test}, 
let $\sigma$ denote the uniform probability measure on the sphere with respect to the Borel sigma algebra $\mathcal{B}$. 
We first observe that the second moments of $\sigma$ satisfy   
\begin{equation*}
M_{i,j}(\sigma) := \int_{S^{d-1}} x^{(i)}x^{(j)}d\sigma(x)=\frac{1}{d}\delta_{i,j},
\end{equation*}
where $x=(x^{(1)},\ldots,x^{(d)})^\top\in\R^d$ and $i,j=1,\ldots,d$. Note that the \emph{Fisher (or scatter) matrix}, 
\begin{equation*}
T_{\{x_i\}_{i=1}^n}=\frac{1}{n}\sum_{i=1}^n x_ix_i^\top,
\end{equation*}
of a sample $\{x_i\}_{i=1}^n\subset S^{d-1}$ equals the matrix of second moments of the underlying counting measure. Recalling that the matrix of second moments of the uniform measure equals $\frac{1}{d}\mathcal{I}_d$, the \emph{Bingham test} rejects the hypothesis of directional uniformity of a sample if its Fisher matrix $T_{\{x_i\}_{i=1}^n}$ is far from $\frac{1}{d}\mathcal{I}_d$. In fact, the Bingham statistic $\frac{d(d+2)}{2}n(\trace(T_{\{x_i\}_{i=1}^n}^2)-\frac{1}{d})$ under uniformity is $\chi^2_{(d-1)(d+2)/2}$ distributed with an error $\mathcal{O}(n^{-1})$, cf.~\cite{Mardia:2008aa}.









We say that the Bingham test is inconsistent when rejection of uniformity fails for a particular non-uniform sample distribution. Here, we focus on those distributions that are multi-modal, where a \emph{mode} is a local maximum of the distribution's density.  
Other analysis tools have been customized to spherical and more general manifold data in \cite{Fletcher:2004aa,Huckemann:2006aa,Jung:2009ab}.



\section{Frames}\label{section:frames}
A collection of points $\{x_i\}_{i=1}^n\subset\R^d$ is called a \emph{finite frame for $\R^d$} if there are two constants $0<A\leq B$, called \emph{lower and upper frame bounds}, respectively, such that
\begin{equation*}
A\|x\|^2 \leq \sum_{i=1}^n |\langle x,x_i\rangle|^2 \leq B\|x\|^2,\quad\text{for all $x\in\R^d$,}
\end{equation*}
where $\langle \cdot , \cdot \rangle$ denotes the usual inner product on $\R^d$. A frame spans $\R^d$, and any finite spanning set is a frame \cite{Christensen:2003aa}. A collection of points $\{x_i\}_{i=1}^n\subset\R^d$ is called a \emph{finite tight frame for $\R^d$} if there is a positive constant $A$, 
such that 
\begin{equation*}
A\|x\|^2 = \sum_{i=1}^n |\langle x,x_i\rangle|^2,\quad\text{for all $x\in\R^d$.}
\end{equation*}
Every finite tight frame gives rise to the expansion
\begin{equation}\label{eq:tight frame parseval}
x = \frac{1}{A}\sum_{i=1}^n \langle x,x_i\rangle x_i,\quad\text{for all $x\in\R^d$,}
\end{equation}
which generalizes the Parseval formula for othonormal bases \cite{Christensen:2003aa}. We define a \emph{finite unit norm tight frame (FNTF)} for $\R^d$ as a tight frame whose elements all have unit norm. According to \cite{Goyal:1998aa}, the tight frame bound $A$ of a FNTF is $n/d$. The collection $\{x_i\}_{i=1}^n\subset S^{d-1}$ is a FNTF iff its Fisher matrix equals $\frac{1}{d}\mathcal{I}_d$ \cite{Christensen:2003aa}. The literature contains many FNTFs. For example, \cite{Sustik:2007aa} shows a FNTF for $\R^3$ composed of $6$ vectors, and $\{(\cos(\alpha_k),\sin(\alpha_k))^\top  : k=1,\ldots,n\}$ is a FNTF for $\R^2$ iff $\sum_{k=1}^n e^{2 i \alpha_k}=0$ \cite{Goyal:2001aa}. In fact, any set of $m$ unit norm vectors in $\R^d$ can be converted to a FNTF by adding $m(d-1)$ extra vectors \cite{Casazza:2003aa}.

We recall probabilistic frames as introduced in \cite{Ehler:2010aa}. Let $\mathcal{M}(\mathcal{B},S^{d-1})$ denote the collection of probability measures on the sphere with respect to the Borel sigma algebra $\mathcal{B}$. An element $\mu\in\mathcal{M}(\mathcal{B},S^{d-1})$ is called a \emph{probabilistic unit norm frame for $\R^d$} if there are constants $0<A\leq B$ such that
 \begin{equation}\label{eq:FNTF prob}
 A\|x\|^2 \leq \int_{S^{d-1}} |\langle x,y\rangle |^2 d\mu (y) \leq B\|x\|^2,\quad\text{for all $x\in\R^d$.}
 \end{equation}
If we can choose $A=B$ in \eqref{eq:FNTF prob}, then we call $\mu$ a \emph{probabilistic unit norm tight frame for $\R^d$}, and $A$ must be equal to $\frac{1}{d}$ \cite{Ehler:2010aa}. We then have
 \begin{equation*}
 x =d \int_{S^{d-1}} \langle x, y\rangle y d\mu(y), \quad\text{for all $x\in\R^d$,}
 \end{equation*}
where the integral is vector valued. This generalizes \eqref{eq:tight frame parseval}, and a sequence of pairwise distinct vectors $\{x_i\}_{i=1}^n\subset S^{d-1}$ is a FNTF for $\R^d$ iff the normalized counting measure $\frac{1}{n}\mu_{x_1\ldots,x_n}$ is a probabilistic unit norm tight frame for $\R^d$.

\section{Joining frame theory and directional statistics}\label{section:joining}

As introduced in Section \ref{section:directional statistics}, the Rayleigh test rejects uniformity if the mean resultant length is far from $0$, while the Bingham test rejects uniformity if the sample's Fisher matrix is far from $\frac{1}{d}\mathcal{I}_d$. We therefore call a probability measure $\mu$ on the sphere a \emph{Rayleigh}-alternative if its mean $\bar{\mu}=\int_{S^{d-1}} x d\mu(x)$ is $0$ and a \emph{Bingham}-alternative if 
$M_{i,j}(\mu)=\frac{1}{d}\delta_{i,j}$ for all $i,j=1\ldots,d$. 
We first characterize Rayleigh- and Bingham-alternatives in terms of maximizers and minimizers, respectively, of certain potentials. Subsequently, we provide a connection to probabilistic frames. 

Bjoerck verifies in \cite{Bjoerck:1955aa} that, among all probability measures $\mu\in\mathcal{M}(\mathcal{B},S^{d-1})$, the maximizers of the \emph{probabilistic Riesz-$2$-potential}
\begin{equation}\label{eq:Riesz potential}
\int_{S^{d-1}} \int_{S^{d-1}}  \|x-y\|^2 d\mu(x)d\mu(y)
\end{equation}
are exactly the zero mean probability measures. Therefore, the maximizers of \eqref{eq:Riesz potential} are the Rayleigh-alternatives.

To characterize Bingham-alternatives, we introduce the \emph{directional force} $F$ between two points $a$ and $b$ on the sphere $S^{d-1}$ as 
$
 F(a,b)=2|\langle a,b\rangle| (a-b).
$
The physical potential between $a$ and $b$ is $|\langle a,b\rangle |^2$ \cite{Benedetto:2003aa}. The minimizers of the \emph{probabilistic frame potential}
\begin{equation}\label{eq:pfp}
\PFP(\mu) = \int_{S^{d-1}}\int_{S^{d-1}} |\langle x,y\rangle|^2 d\mu(x)d\mu(y),
\end{equation}
among all probability measures $\mu\in\mathcal{M}(\mathcal{B},S^{d-1})$, 
are said to be in \emph{equilibrium under the directional force}, or simply in \emph{directional equilibrium} \cite{Benedetto:2003aa,Ehler:2010aa}. Naturally, the uniform distribution is in directional equilibrium \cite{Ehler:2010aa}; however, other distributions with a mixture of well-defined modes can also be in directional equilibrium. 
In fact, the minimizers of the directional potential \eqref{eq:pfp} are exactly those probability measures $\mu\in\mathcal{M}(\mathcal{B},S^{d-1})$ whose second moments satisfy $M_{i,j}(\mu)=\delta_{i,j}\frac{1}{d}$ \cite{Ehler:2010aa} and are, therefore, the Bingham-alternatives. These minimizers were characterized as the probabilistic unit norm tight frames for $\R^d$ in \cite{Ehler:2010aa}. 
The advantage of the latter characterization is that tight frames are well-understood in terms of algebraic as well as geometric conditions \cite{Benedetto:2003aa,Christensen:2003aa,Ehler:2010aa}.


Results in \cite{Ehler:2010aa} and the present work imply that the Bingham-alternatives with zero mean are the minimizers of the \emph{fractional frame-Riesz-$2$-potential}
\begin{equation}\label{eq:mixed potential}
\frac{\int_{S^{d-1}} \int_{S^{d-1}} |\langle x,y\rangle |^2 d\mu(x) d\mu(y)}{\int_{S^{d-1}} \int_{S^{d-1}}  \|x-y\|^2 d\mu(x)d\mu(y)}.
\end{equation}
Hence, probabilistic unit norm tight frames with zero mean are both, Rayleigh- and Bingham-alternatives.

\section{Patterning of granular rods}
\subsection{Order-disorder phase transition}
A collection of rod shaped particles can undergo an order-disorder phase transition 
that, in the simplest model, is controlled by entropy. At low rod densities, the maximal total entropy occurs when both rotational and translational entropies are independently maximized. Beyond a critical density, however, 
rotational entropy is sacrificed for significant gains in translational entropy, resulting in a phase transition from randomly (uniformly) rotated rods, Fig.~1(a), to directionally oriented rods, Fig.~1(b). A PCA-type method measures the average rod direction (director) and strength of rod alignment (order parameter) that results from rotational entropy loss and is described in the following: 
Let $x_i\in S^{d-1}$ denote the direction of the $i$-th rod out of $n$ total rods. For simplicity, let us assume that the alignment is measured in a plane, hence $d=2$. The covariance type matrix  
\begin{equation}\label{eq:PCA directions}
Q_2 =  \frac{1}{n}\sum_{i=1}^n 2 x_i x_i^\top -\mathcal{I}_2
\end{equation}
is therefore used to determine the director. Since $Q_2$ is symmetric, the eigenvectors form an orthogonal basis, where the nonnegative eigenvalue $\lambda$ corresponds to the order parameter and the associated eigenvector corresponds to the director, cf.~\cite{Frenkel:1985aa,Galanis:2006aa}. In fact, $0\leq \lambda\leq 1$ and the second eigenvalue of $Q_2$ equals $-\lambda$. 

This PCA-type method only measures unidirectional rod ordering. In experiments, however, fluctuations and/or competing forces like container boundaries, Figs.~1(c) and 1(d), can influence rod alignment. Therefore, the director field can vary spatially, potentially resulting in complex multidirectional patterning, Fig.~1(d), that sometimes cannot be distinguished from a disordered state when analyzed by the traditional order parameter. 




 \begin{figure}[h]
\subfigure[]{\includegraphics[height=.22\textwidth]{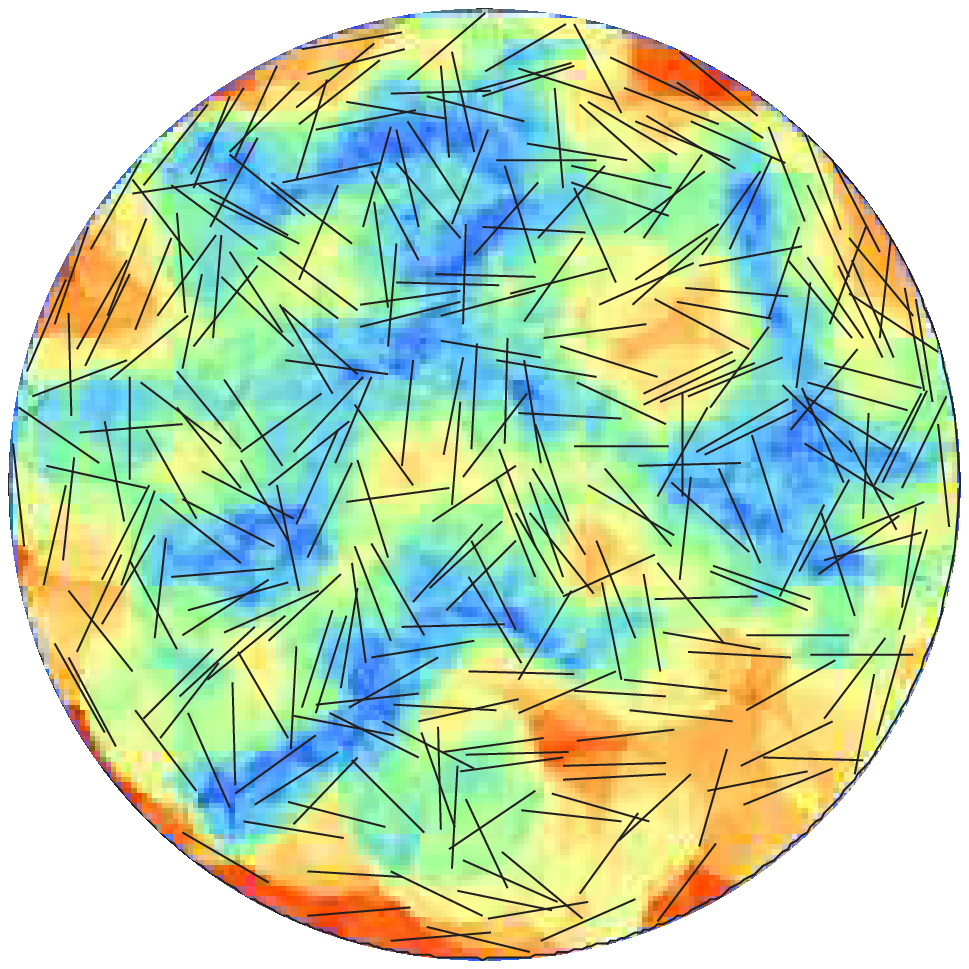}}
\label{subfigure:a}
\subfigure[]{\includegraphics[height=.22\textwidth]{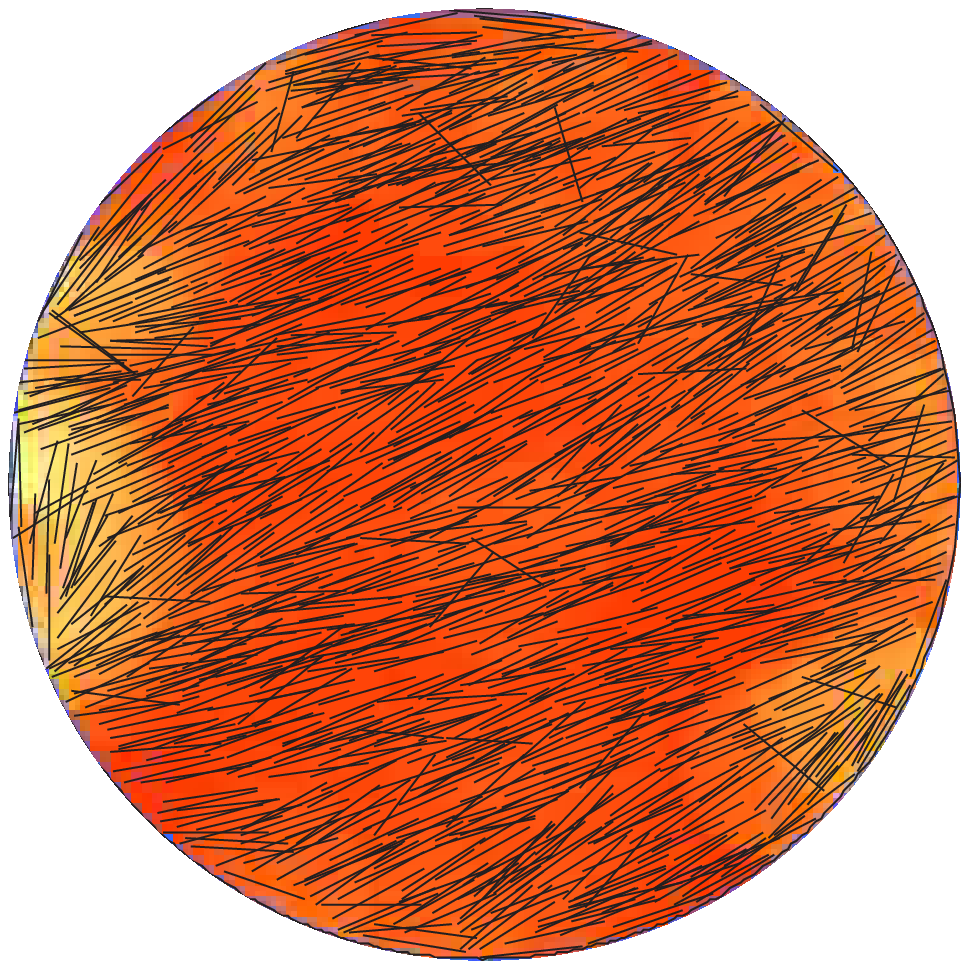}}
\subfigure[]{
\includegraphics[height=.22\textwidth]{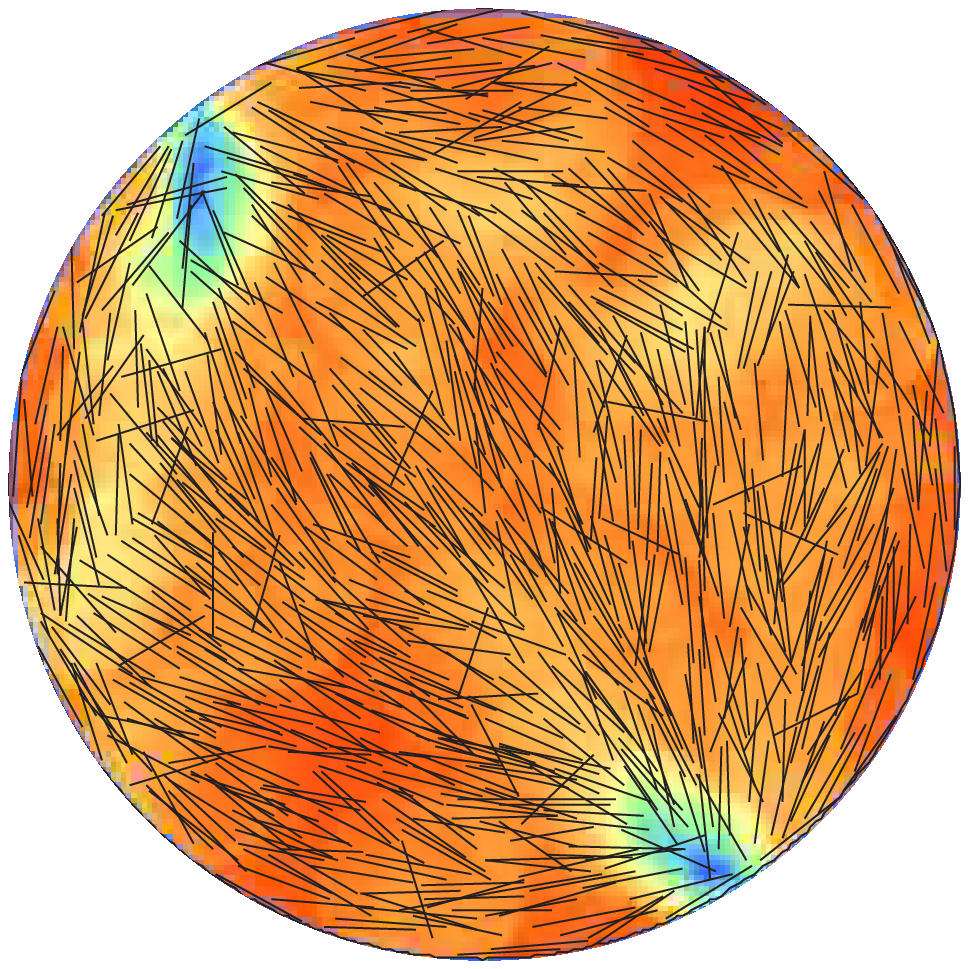}}
\subfigure[ \hspace{.5cm}]{\includegraphics[height=.22\textwidth]{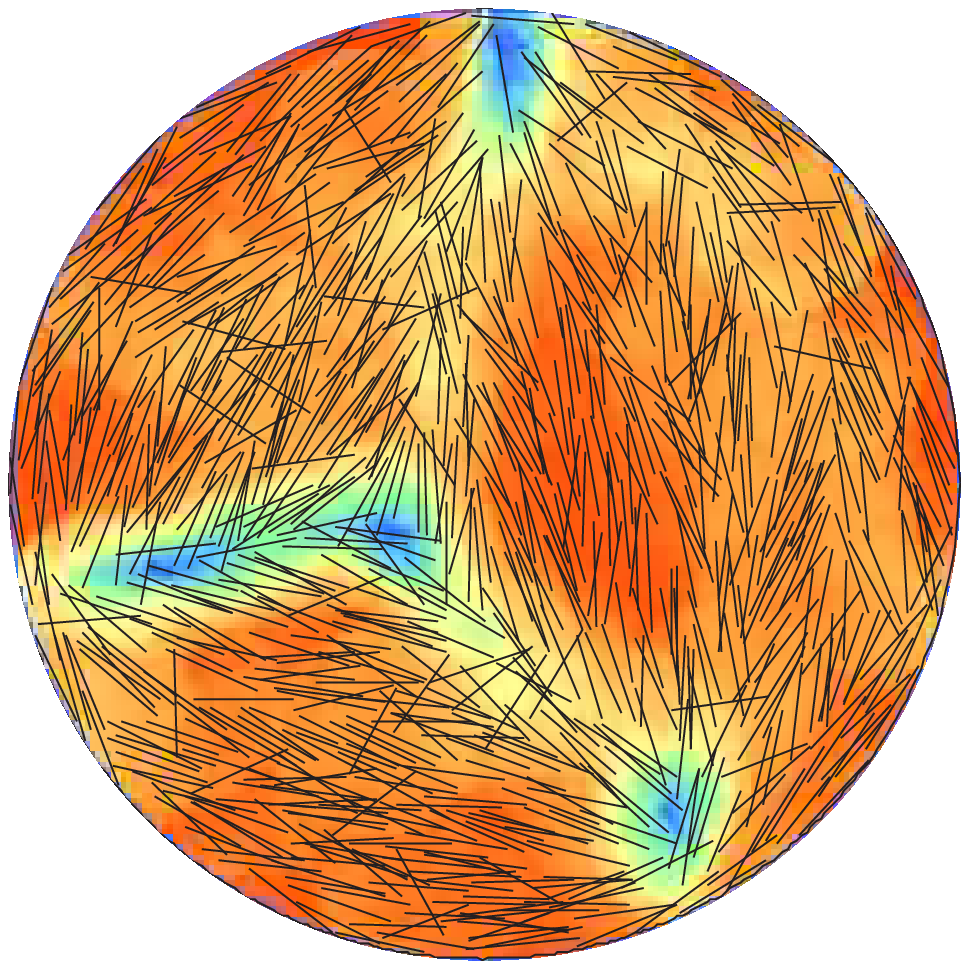}\includegraphics[height=.22\textwidth]{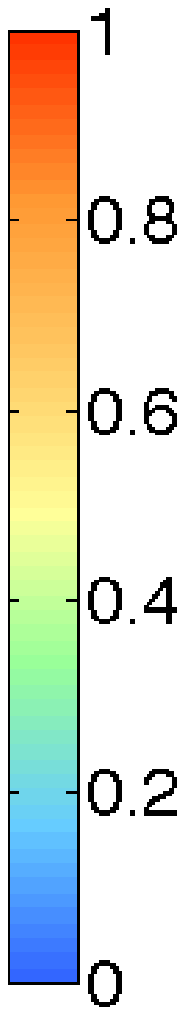}
}

\includegraphics[height=.22\textwidth]{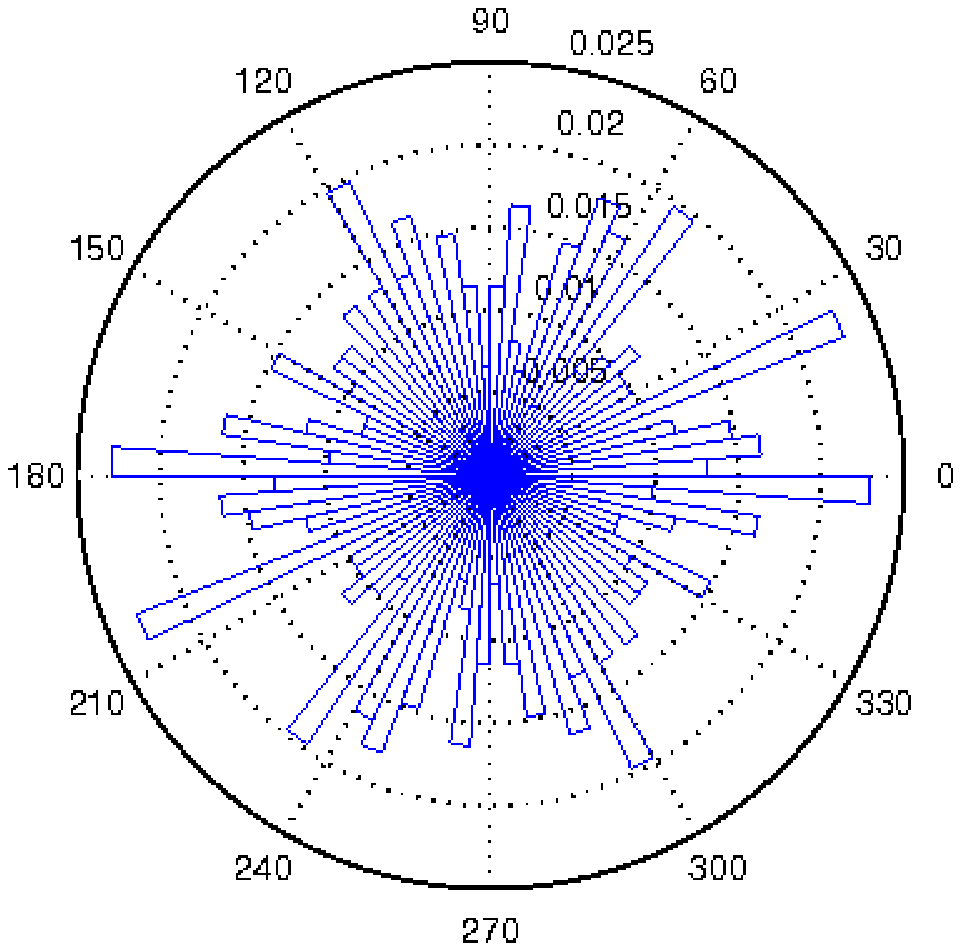}
\includegraphics[height=.22\textwidth]{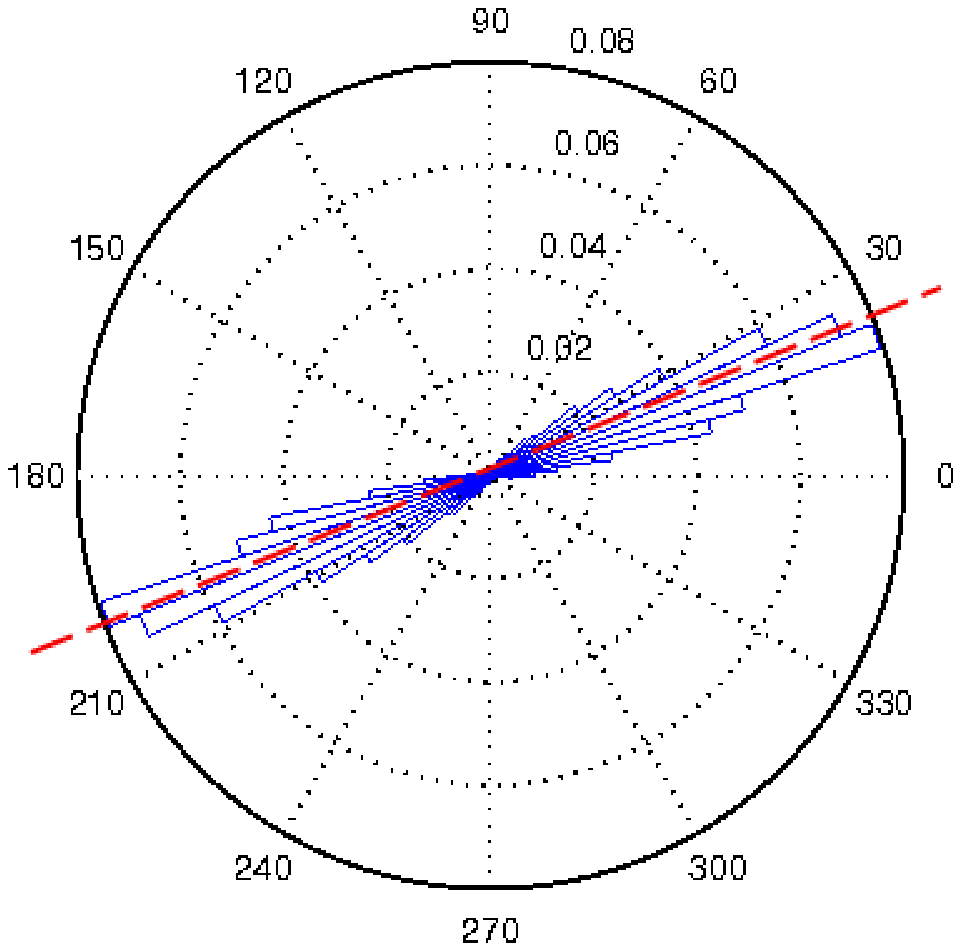}
\includegraphics[height=.22\textwidth]{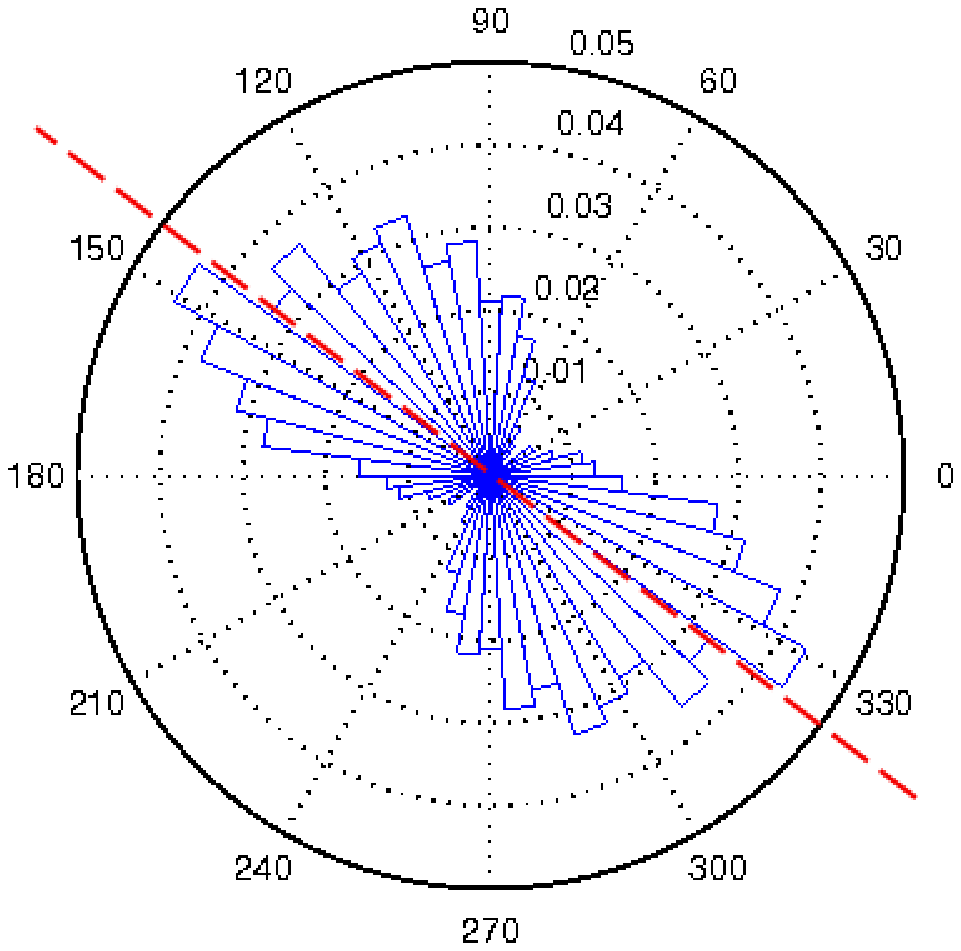}
\includegraphics[height=.22\textwidth]{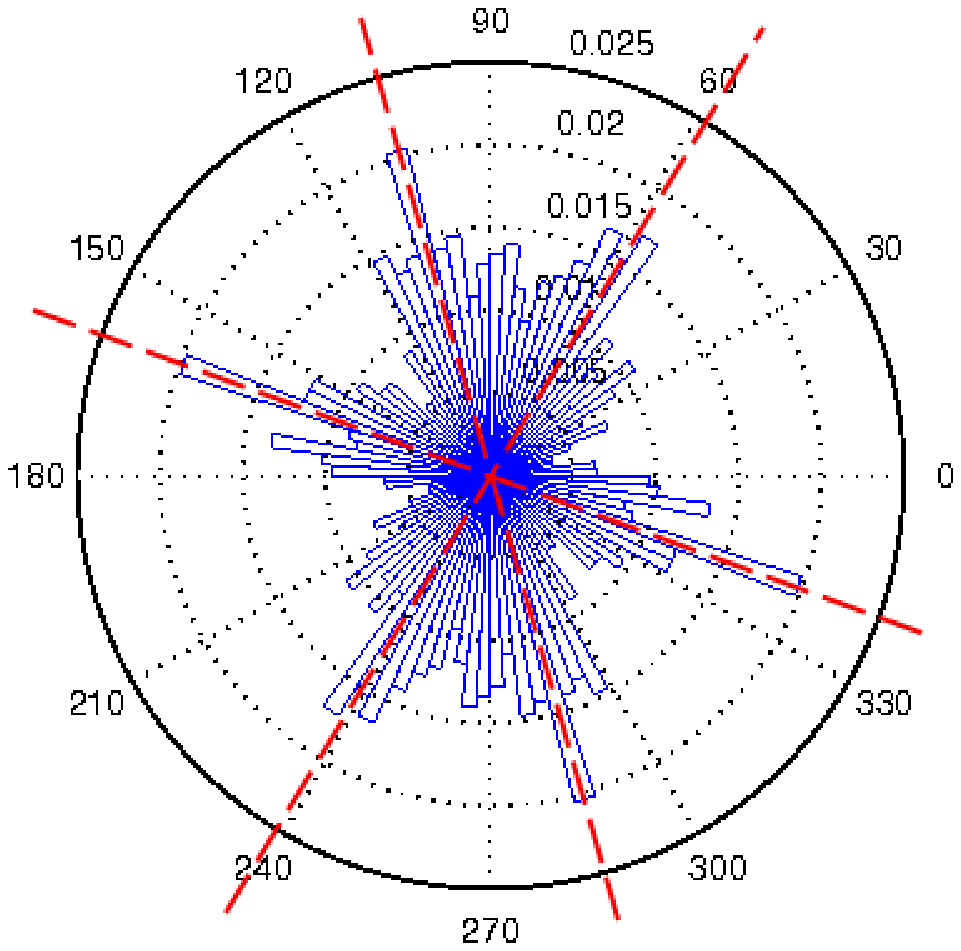}

\hspace{.22\textwidth}
\includegraphics[height=.22\textwidth]{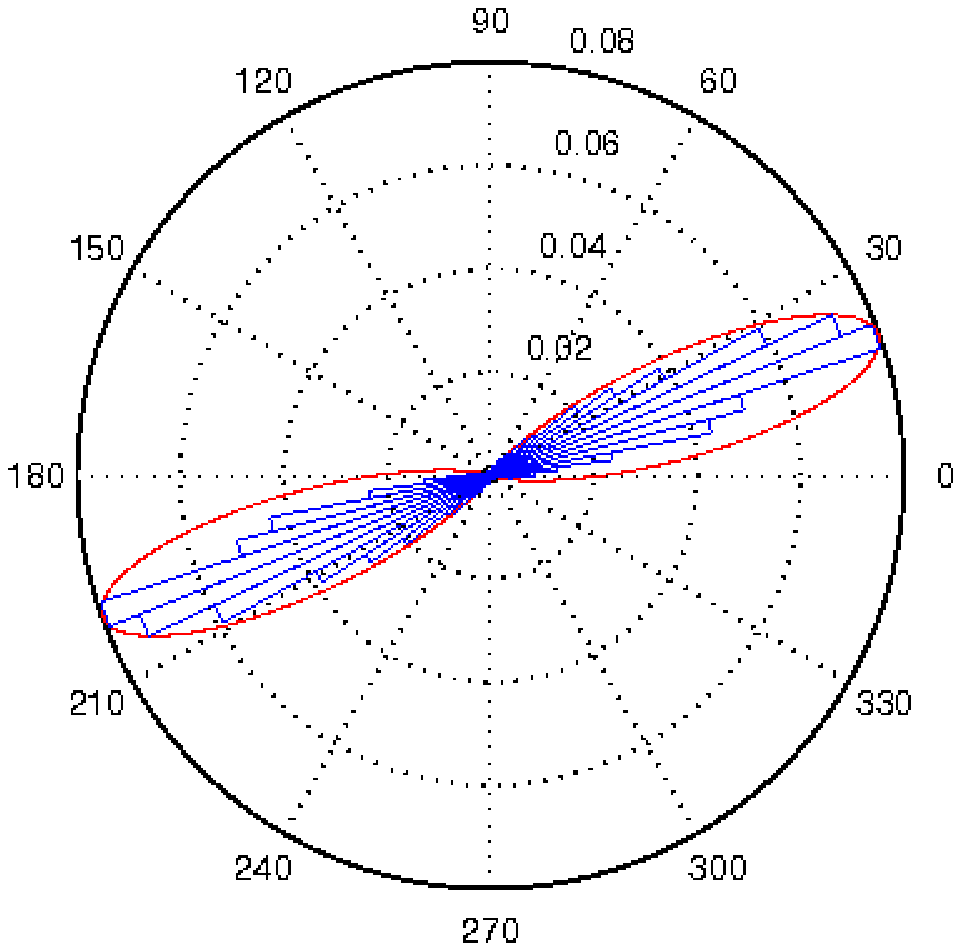}
\includegraphics[height=.22\textwidth]{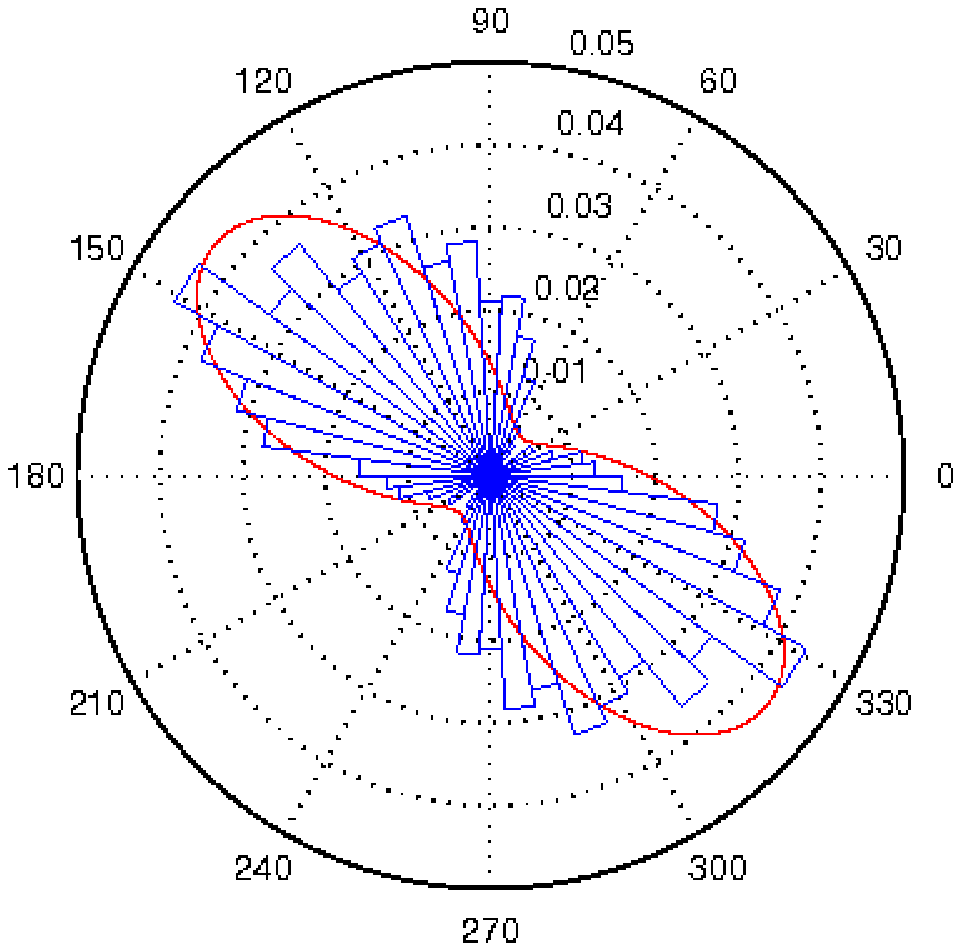}
\includegraphics[height=.22\textwidth]{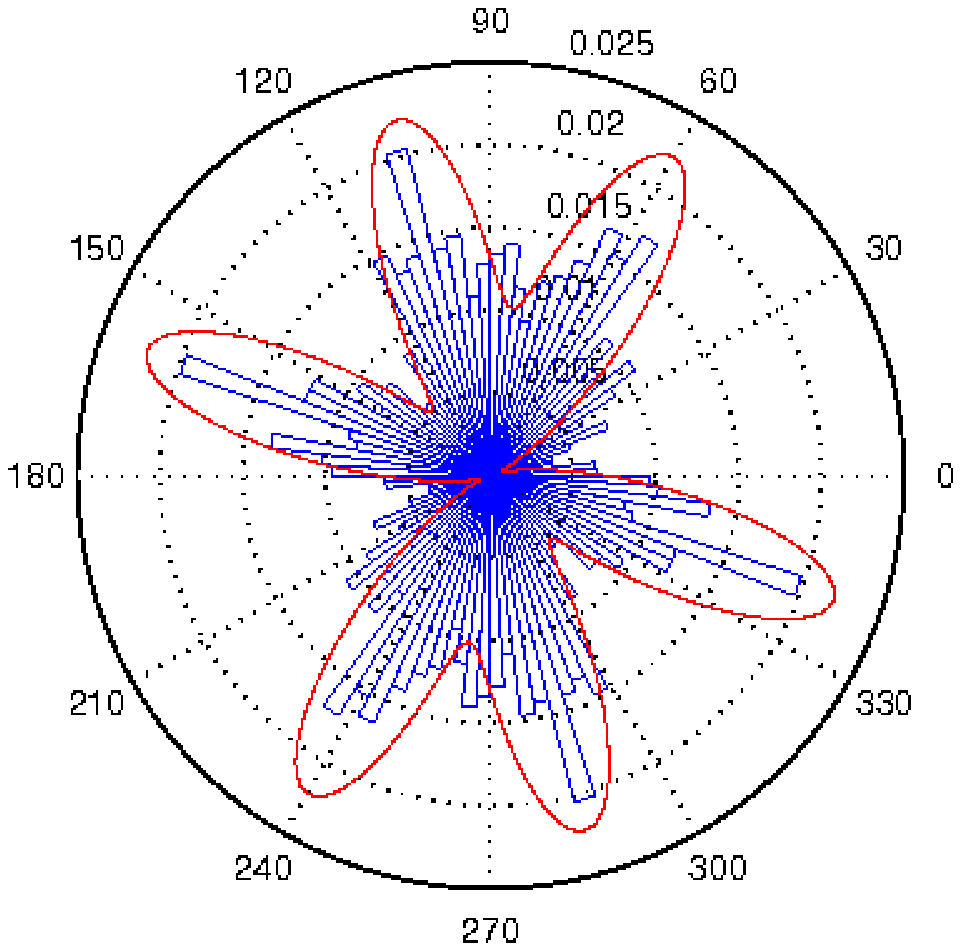}
\caption{Rod patterns observed in 2D. (top) Images from granular rod experiments showing a disordered state (a), unidirectional ordered states with small (b) and large (c) standard deviations, and a multidirectional ordered state (d). The order parameters derived from $Q_2$ in \eqref{eq:PCA directions} are $0.0950$ (a), $0.8902$ (b), $0.4267$ (c), and $0.1534$ (d). The Bingham test rejects uniformity at a $1\%$ confidence level for (b) and (c); however, uniformity cannot be rejected at a $5\%$ confidence level for (a) and (d).  
Color represents the local order parameter, where $Q_2$ is measured only in a vicinity of less than one rod length from each location in the image. (center) Directional histograms (blue) and directors $z_i$ (dashed red lines) associated with each state. (bottom) We apply the Watson  mixture model in \eqref{eq:mix von Mises in prop} with a suitable $\kappa$.  
}\label{figure:figure}
\end{figure}

\subsection{New model by means of directional statistics}\label{section:the end}
We propose a complementary analysis of rod ordering that can more accurately quantify multidirectional alignment. We fit a probability model to experimentally observed rod patterning by first identifying a set of directions, cf.~Fig.~1(b)-(d). 
Since a rod rotated by $180$ degrees is indistinguishable from an unrotated rod, we do not differentiate between $x$ and $-x$. Consistent with this criteria,  
let $\sigma\in \mathcal{M}(\mathcal{B},S^{d-1})$ be the uniform measure on the sphere, $z_0\in S^{d-1}$, and $\kappa\neq 0$, the \emph{Watson measure} $\mu$ is then given by
\begin{equation}\label{eq:Mises-Fisher}
\mu(x) = c_d(\kappa) \exp(\kappa \langle z_0, x\rangle^2)\sigma(x),
\end{equation}
where $c_d(\kappa)=\frac{\Gamma(d/2)}{2\pi^{d/2} F(1/2,d/2,\kappa)}$, $\Gamma$ the usual Gamma function, and $F$ a confluent hypergeometric function \cite{Mardia:2008aa}. For $\kappa > 0$, the density tends
to concentrate around $\pm z_0$, whereas for $\kappa < 0$, the density concentrates around the great circle orthogonal to $z_0$. And as  $|\kappa|$ increases, the density peaks tighten.

Next, we model each sample with a mixture of Watson distributions, i.e., for a collection of directors $\{z_i\}_{i=1}^N\subset S^1$, we consider
\begin{equation}\label{eq:mix von Mises in prop}
\mu(x) = \frac{c_2(\kappa)}{N}\sum_{i=1}^N  \exp(\kappa \langle z_i , x\rangle^2) \sigma(x).
\end{equation}
By replacing von Mises measures in  \cite{Ehler:2010aa} with Watson measures, we conclude the following: 
if $\{z_i\}_{i=1}^N$ is a FNTF for $\R^2$, then, for any $\kappa\neq 0$, the measure \eqref{eq:mix von Mises in prop} is a probabilistic unit norm tight frame for $\R^2$, hence a Bingham-alternative.  Note that in $\R^2$, a FNTF with three elements (directions) must be equiangular, and all equiangular tri-directions are FNTFs \cite{Goyal:2001aa}. Nevertheless, if a sample is distributed so that the modes in \eqref{eq:mix von Mises in prop} approximate a minimizer, like the three nearly equiangular directions in Fig.~1(d), then the Bingham test may also fail to reject uniformity, as is the case with this figure.

Finally, this approach extracts two parameter sets from the sample, directional modes and associated widths \cite{Bartels:1984aa,Fisher:1995aa,HSU:1986aa}, where the widths represent a measure for directional ordering. These parameters will be used in a forthcoming paper to quantify the differences between experimental rod patterns and the expected behavior from theories and simulations. Furthermore, these tools may provide a method to identify more subtle rod patterning transitions, like the one described in \cite{Galanis:2010ab}.

\section*{Acknowledgements}
The authors are supported, in part, by intramural research funds from the National Institute of Child Health and Human Development, National Institutes of Health. The Fritz Haber research center is supported by the Minerva Foundation, Munich, Germany. In addition, ME is supported by the
NIH/DFG Research Career Transition Awards Program (EH 405/1-1/575910).







\begin{thebibliography}{36}
\expandafter\ifx\csname natexlab\endcsname\relax\def\natexlab#1{#1}\fi
\providecommand{\bibinfo}[2]{#2}
\ifx\xfnm\relax \def\xfnm[#1]{\unskip,\space#1}\fi
\bibitem[{Bartels(1984)}]{Bartels:1984aa}
\bibinfo{author}{R.~Bartels}, \bibinfo{title}{Estimation in a bidirectional
  mixture of von mises distributions}, \bibinfo{journal}{Biometrics}
  \bibinfo{volume}{40} (\bibinfo{year}{1984}) \bibinfo{pages}{777--784}.
\bibitem[{Benedetto and Fickus(2003)}]{Benedetto:2003aa}
\bibinfo{author}{J.J. Benedetto}, \bibinfo{author}{M.~Fickus},
  \bibinfo{title}{Finite normalized tight frames}, \bibinfo{journal}{Adv.~
  Comput.~ Math.} \bibinfo{volume}{18} (\bibinfo{year}{2003})
  \bibinfo{pages}{357--385}.
\bibitem[{Bjoerck(1955)}]{Bjoerck:1955aa}
\bibinfo{author}{G.~Bjoerck}, \bibinfo{title}{Distributions of positive mass,
  which maximize a certain generalized energy integral},
  \bibinfo{journal}{Arkiv foer Matematik} \bibinfo{volume}{3}
  (\bibinfo{year}{1955}) \bibinfo{pages}{255--269}.
\bibitem[{Boomsma et~al.(2008)Boomsma, Mardia, Taylor, Ferkinghoff-Borg, Krogh
  and Hamelryck}]{Boomsma:2008aa}
\bibinfo{author}{W.~Boomsma}, \bibinfo{author}{K.V. Mardia},
  \bibinfo{author}{C.C. Taylor}, \bibinfo{author}{J.~Ferkinghoff-Borg},
  \bibinfo{author}{A.~Krogh}, \bibinfo{author}{T.~Hamelryck}, \bibinfo{title}{A
  generative, probabilistic model of local protein structure},
  \bibinfo{journal}{Proc.~ Nat.~ Acad.~ Sci.} \bibinfo{volume}{105}
  (\bibinfo{year}{2008}) \bibinfo{pages}{8932--8937}.
\bibitem[{Casazza and Kovacevic(2003)}]{Casazza:2003aa}
\bibinfo{author}{P.G. Casazza}, \bibinfo{author}{J.~Kovacevic},
  \bibinfo{title}{Equal-norm tight frames with erasures},
  \bibinfo{journal}{Adv.~ Comput.~ Math.} \bibinfo{volume}{18}
  (\bibinfo{year}{2003}) \bibinfo{pages}{387--430}.
\bibitem[{Chang et~al.(2010)Chang, Ha, Slaughter and Link}]{Chang:2010aa}
\bibinfo{author}{W.~Chang}, \bibinfo{author}{J.W. Ha}, \bibinfo{author}{L.S.
  Slaughter}, \bibinfo{author}{S.~Link}, \bibinfo{title}{Plasmonic nanorod
  absorbers as orientation sensors}, \bibinfo{journal}{Proc.~ Nat.~ Acad.~
  Sci.} \bibinfo{volume}{107} (\bibinfo{year}{2010})
  \bibinfo{pages}{2781--2786}.
\bibitem[{Chass{\'e} and Th{\'e}ron(1988)}]{Chasse:1988aa}
\bibinfo{author}{J.L. Chass{\'e}}, \bibinfo{author}{A.~Th{\'e}ron},
  \bibinfo{title}{An example of circular statistics in chronobiological
  studies: analysis of polymorphism in the emergence rhythms of schistosoma
  mansoni cercariae}, \bibinfo{journal}{Chronobiol.~ Int.} \bibinfo{volume}{5}
  (\bibinfo{year}{1988}) \bibinfo{pages}{433--9}.
\bibitem[{Christensen(2003)}]{Christensen:2003aa}
\bibinfo{author}{O.~Christensen}, \bibinfo{title}{{A}n {I}ntroduction to
  {F}rames and {R}iesz {B}ases}, \bibinfo{publisher}{Birkh{\"{a}}user},
  \bibinfo{address}{Boston}, \bibinfo{year}{2003}.
\bibitem[{Daubechies et~al.(1986)Daubechies, Grossmann and
  Meyer}]{Daubechies:1986aa}
\bibinfo{author}{I.~Daubechies}, \bibinfo{author}{A.~Grossmann},
  \bibinfo{author}{Y.~Meyer}, \bibinfo{title}{Painless nonorthogonal
  expansions}, \bibinfo{journal}{J.~ Math.~ Phys.} \bibinfo{volume}{27}
  (\bibinfo{year}{1986}) \bibinfo{pages}{1271--1283}.
\bibitem[{Ehler(2007)}]{Ehler:2007aa}
\bibinfo{author}{M.~Ehler}, \bibinfo{title}{On multivariate compactly supported
  bi-frames}, \bibinfo{journal}{J.~ Fourier Anal.~ Appl.} \bibinfo{volume}{13}
  (\bibinfo{year}{2007}) \bibinfo{pages}{511--532}.
\bibitem[{Ehler(2009)}]{Ehler:ab}
\bibinfo{author}{M.~Ehler}, \bibinfo{title}{Nonlinear approximation associated
  with nonseparable wavelet bi-frames}, \bibinfo{journal}{J.~ Approx.~ Theory}
  \bibinfo{volume}{161} (\bibinfo{year}{2009}) \bibinfo{pages}{292--313}.
\bibitem[{Ehler(2010{\natexlab{a}})}]{Ehler:2010ae}
\bibinfo{author}{M.~Ehler}, \bibinfo{title}{The multiresolution structure of
  pairs of dual wavelet frames for a pair of {S}obolev spaces},
  \bibinfo{journal}{Jaen J.~ Approx.} \bibinfo{volume}{2}
  (\bibinfo{year}{2010}{\natexlab{a}}).
\bibitem[{Ehler(2010{\natexlab{b}})}]{Ehler:2010aa}
\bibinfo{author}{M.~Ehler}, \bibinfo{title}{Random tight frames},
  \bibinfo{journal}{submitted to J.~ Fourier Anal.~ Appl.}
  (\bibinfo{year}{2010}{\natexlab{b}}).
\bibitem[{Ehler and Han(2008)}]{Ehler:2008ab}
\bibinfo{author}{M.~Ehler}, \bibinfo{author}{B.~Han}, \bibinfo{title}{Wavelet
  bi-frames with few generators from multivariate refinable functions},
  \bibinfo{journal}{Appl.~ Comput.~ Harmon.~ Anal.} \bibinfo{volume}{25}
  (\bibinfo{year}{2008}) \bibinfo{pages}{407--414}.
\bibitem[{Ehler and Koch(2010)}]{Ehler:aa}
\bibinfo{author}{M.~Ehler}, \bibinfo{author}{K.~Koch}, \bibinfo{title}{The
  construction of multiwavelet bi-frames and applications to variational image
  denoising}, \bibinfo{journal}{Int.~ J.~ Wavelets, Multiresolut.~ Inf.~
  Process.} \bibinfo{volume}{8} (\bibinfo{year}{2010})
  \bibinfo{pages}{431--455}.
\bibitem[{Feichtinger and Strohmer(2003)}]{Feichtinger:2003aa}
\bibinfo{author}{H.G. Feichtinger}, \bibinfo{author}{T.~Strohmer},
  \bibinfo{title}{{A}dvances in {G}abor {A}nalysis},
  \bibinfo{publisher}{Birkh{\"{a}}user}, \bibinfo{address}{Boston},
  \bibinfo{year}{2003}.
\bibitem[{Fisher(1995)}]{Fisher:1995aa}
\bibinfo{author}{N.I. Fisher}, \bibinfo{title}{Statistical Analysis of Circular
  Data}, \bibinfo{publisher}{Cambridge University Press}, \bibinfo{year}{1995}.
\bibitem[{Fletcher et~al.(2004)Fletcher, Lu, Pizer and Joshi}]{Fletcher:2004aa}
\bibinfo{author}{P.T. Fletcher}, \bibinfo{author}{C.~Lu}, \bibinfo{author}{S.M.
  Pizer}, \bibinfo{author}{S.~Joshi}, \bibinfo{title}{Principal geodesic
  analysis for the study of nonlinear statistics of shape},
  \bibinfo{journal}{IEEE Trans.~ Med.~ Imaging} \bibinfo{volume}{23}
  (\bibinfo{year}{2004}) \bibinfo{pages}{995--1005}.
\bibitem[{Frenkel and Eppenga(1985)}]{Frenkel:1985aa}
\bibinfo{author}{D.~Frenkel}, \bibinfo{author}{R.~Eppenga},
  \bibinfo{title}{Evidence for algebraic orientational order in a
  two-dimensional hard-core nematic}, \bibinfo{journal}{Phys.~ Rev.~ A.}
  \bibinfo{volume}{31} (\bibinfo{year}{1985}) \bibinfo{pages}{1776--1787}.
\bibitem[{Galanis et~al.(2006)Galanis, Harries, Sackett, Losert and
  Nossal}]{Galanis:2006aa}
\bibinfo{author}{J.~Galanis}, \bibinfo{author}{D.~Harries},
  \bibinfo{author}{D.L. Sackett}, \bibinfo{author}{W.~Losert},
  \bibinfo{author}{R.~Nossal}, \bibinfo{title}{Spontaneous patterning of
  confined granular rods}, \bibinfo{journal}{Phys.~ Rev.~ Lett.}
  \bibinfo{volume}{96} (\bibinfo{year}{2006}) \bibinfo{pages}{028002}.
\bibitem[{Galanis et~al.(2010)Galanis, Nossal, Losert and
  Harries}]{Galanis:2010ab}
\bibinfo{author}{J.~Galanis}, \bibinfo{author}{R.~Nossal},
  \bibinfo{author}{W.~Losert}, \bibinfo{author}{D.~Harries},
  \bibinfo{title}{Nematic order in small systems: measuring the elastic and
  wall-anchoring constants in vibrofluidized granular rods},
  \bibinfo{journal}{Phys.~ Rev.~ Lett.} \bibinfo{volume}{105}
  (\bibinfo{year}{2010}).
\bibitem[{de~Gennes and Prost(1993)}]{Gennes:1993aa}
\bibinfo{author}{P.G. de~Gennes}, \bibinfo{author}{J.~Prost},
  \bibinfo{title}{The Physics of Liquid Crystals}, \bibinfo{publisher}{Oxford
  Science}, \bibinfo{address}{New York}, \bibinfo{year}{1993}.
\bibitem[{Goyal et~al.(2001)Goyal, Kovacevic and Kelner}]{Goyal:2001aa}
\bibinfo{author}{V.K. Goyal}, \bibinfo{author}{J.~Kovacevic},
  \bibinfo{author}{J.A. Kelner}, \bibinfo{title}{Quantized frame expansions
  with erasures}, \bibinfo{journal}{Appl.~ Comput.~ Harmon.~ Anal.}
  \bibinfo{volume}{10} (\bibinfo{year}{2001}) \bibinfo{pages}{203--233}.
\bibitem[{Goyal et~al.(1998)Goyal, Vetterli and Thao}]{Goyal:1998aa}
\bibinfo{author}{V.K. Goyal}, \bibinfo{author}{M.~Vetterli},
  \bibinfo{author}{N.T. Thao}, \bibinfo{title}{Quantized overcomplete
  expansions in $\mathbb{R}^n$: Analysis, synthesis, and algorithms},
  \bibinfo{journal}{IEEE Trans.~ Inform.~ Theory} \bibinfo{volume}{44}
  (\bibinfo{year}{1998}) \bibinfo{pages}{16--31}.
\bibitem[{Gr{\"{o}}chenig(2001)}]{Grochenig:2001aa}
\bibinfo{author}{K.~Gr{\"{o}}chenig}, \bibinfo{title}{{F}oundations of
  {T}ime-{F}requency {A}nalysis}, \bibinfo{publisher}{Birkh{\"{a}}user},
  \bibinfo{address}{Boston}, \bibinfo{year}{2001}.
\bibitem[{Hsu et~al.(1986)Hsu, Walker and Ogren}]{HSU:1986aa}
\bibinfo{author}{Y.S. Hsu}, \bibinfo{author}{J.J. Walker},
  \bibinfo{author}{D.E. Ogren}, \bibinfo{title}{A stepwise method for
  determining the number of component distributions in a mixture},
  \bibinfo{journal}{Math.~ Geol.} \bibinfo{volume}{18} (\bibinfo{year}{1986})
  \bibinfo{pages}{153--160}.
\bibitem[{Huckemann and Ziezold(2006)}]{Huckemann:2006aa}
\bibinfo{author}{S.~Huckemann}, \bibinfo{author}{H.~Ziezold},
  \bibinfo{title}{Principal component analysis for riemannian manifolds, with
  an application to triangular shape spaces}, \bibinfo{journal}{Adv.~ in Appl.~
  Probab.} \bibinfo{volume}{38} (\bibinfo{year}{2006})
  \bibinfo{pages}{299--319}.
\bibitem[{Hudson and Thomas(1989)}]{Hudson:1989aa}
\bibinfo{author}{S.D. Hudson}, \bibinfo{author}{E.L. Thomas},
  \bibinfo{title}{Frank elastic-constant anisotropy measured from
  transmission-electron-microscope images of disclinations},
  \bibinfo{journal}{Phys.~ Rev.~ Lett.} \bibinfo{volume}{62}
  (\bibinfo{year}{1989}) \bibinfo{pages}{1993--1996}.
\bibitem[{Jung et~al.(2010)Jung, Foskey and Marron}]{Jung:2009ab}
\bibinfo{author}{S.~Jung}, \bibinfo{author}{M.~Foskey}, \bibinfo{author}{J.S.
  Marron}, \bibinfo{title}{Principal arc analysis on direct product manifolds},
  \bibinfo{journal}{Ann.~ Appl.~ Stat.} \bibinfo{volume}{in press}
  (\bibinfo{year}{2010}).
\bibitem[{Levine et~al.(2002)Levine, Funes, Dowse and Hall}]{Levine:2002aa}
\bibinfo{author}{J.D. Levine}, \bibinfo{author}{P.~Funes},
  \bibinfo{author}{H.B. Dowse}, \bibinfo{author}{J.C. Hall},
  \bibinfo{title}{Resetting the circadian clock by social experience in
  drosophila melanogaster}, \bibinfo{journal}{Science} \bibinfo{volume}{298}
  (\bibinfo{year}{2002}) \bibinfo{pages}{2010--2}.
\bibitem[{Mardia and Jupp(2008)}]{Mardia:2008aa}
\bibinfo{author}{K.V. Mardia}, \bibinfo{author}{P.E. Jupp},
  \bibinfo{title}{Directional Statistics}, Wiley Series in Probability and
  Statistics, \bibinfo{publisher}{John Wiley \& Sons}, \bibinfo{year}{2008}.
\bibitem[{Mardia and Taylor(2007)}]{Mardia:2007aa}
\bibinfo{author}{K.V. Mardia}, \bibinfo{author}{C.C. Taylor},
  \bibinfo{title}{Protein bioinformatics and mixtures of bivariate von mises
  distributions for angular data}, \bibinfo{journal}{Biometrics}
  \bibinfo{volume}{63} (\bibinfo{year}{2007}) \bibinfo{pages}{505--512}.
\bibitem[{Onsager(1949)}]{Onsager:1949aa}
\bibinfo{author}{L.~Onsager}, \bibinfo{title}{The effects of shape on the
  interaction of colloidal particles}, \bibinfo{journal}{Ann.~ N.~ Y.~ Acad.~
  Sci.} \bibinfo{volume}{51} (\bibinfo{year}{1949}) \bibinfo{pages}{627--659}.
\bibitem[{Rueda et~al.(2009)Rueda, Fernandez and Peddada}]{Rueda:2009aa}
\bibinfo{author}{C.~Rueda}, \bibinfo{author}{M.A. Fernandez},
  \bibinfo{author}{S.D. Peddada}, \bibinfo{title}{Estimation of parameters
  subject to order restrictions on a circle with application to estimation of
  phase angles of cell cycle genes}, \bibinfo{journal}{J.~Am.~Stat.~Assoc.}
  \bibinfo{volume}{104} (\bibinfo{year}{2009}) \bibinfo{pages}{338--347}.
\bibitem[{Sustik et~al.(2007)Sustik, Tropp, Dhillon and Heath}]{Sustik:2007aa}
\bibinfo{author}{A.~Sustik}, \bibinfo{author}{J.A. Tropp},
  \bibinfo{author}{I.S. Dhillon}, \bibinfo{author}{R.W. Heath},
  \bibinfo{title}{On the existence of equiangular tight frames},
  \bibinfo{journal}{Linear Algebra Appl.} \bibinfo{volume}{426}
  (\bibinfo{year}{2007}) \bibinfo{pages}{619--635}.
\bibitem[{Xiao et~al.(2010)Xiao, Qiao, He and Yeung}]{Xiao:2010aa}
\bibinfo{author}{L.~Xiao}, \bibinfo{author}{Y.~Qiao}, \bibinfo{author}{Y.~He},
  \bibinfo{author}{E.S. Yeung}, \bibinfo{title}{Three dimensional orientational
  imaging of nanoparticles with darkfield microscopy}, \bibinfo{journal}{Anal.~
  Chem.} \bibinfo{volume}{82} (\bibinfo{year}{2010})
  \bibinfo{pages}{5268--5274}.

\end{thebibliography}







\end{document}